\UseRawInputEncoding
\documentclass[a4paper,12pt]{article}

\def\ad   {a^{\dagger}}

\def\HP   {\hat {P}}

\def\al   {\alpha}

\def\HH {\hat H}

\def\HP {\hat P}

\def\HH {\hat H}

\def\oc {\overline c}

{\it}
{\it}
{\it}
{\it}
{\it}
{\it}

\def\om {\omega}

\usepackage{times}
\usepackage{graphicx}
\begin{document}
\title{  A study of open shell nuclei using  chiral two-body
interactions.}
\author{G. Puddu\\
       Dipartimento di Fisica dell'Universita' di Milano,\\
       Via Celoria 16, I-20133 Milano, Italy}
\maketitle
\begin {abstract}
      We apply the Hybrid-Multi-Determinant method using the recent chiral
      two-body interactions of Entem-Machleidt-Nosyk (EMN) without renormalization to few nuclei
      up to A=48. 
      Mostly we use the bare fifth order NN interaction N4LO-450. 
      For ${}^{24}Mg$ and ${}^{48}Cr$ the excitation energies of the $2^+_1$ states are far larger
      than the corresponding experimental values.
\par\noindent
{\bf{Pacs numbers}}: 21.10.-k,21.60.De
\par\noindent
\vfill
\eject
\end{abstract}
\section{ Introduction.}
\bigskip
\par
      In the past several years we have witnessed the development of powerful ab-initio many-body
      techniques to solve the nuclear Schroedinger equation. Among these methods we mention
      the no-core shell model (NCSM) (ref.[1]), the coupled-cluster (CC) method (ref.[2]) 
      and the in-medium similarity
      renormalization group (IM-SRG) (ref.[3]).
      While the NCSM can only be used for light nuclei because of the exponential increase of the size
      of the Hilbert space with the particle number, for closed shells or around shell closure the
      CC method has been used up to  medium mass nuclei. Quite recently, advances in the Multi-Reference
      IM-SRG (MR-IM-SRG) have been applied to doubly open shell medium mass nuclei (ref.[4]).
      Both CC method and (MR)-IM-SRG scale polynomially with the size of the single-particle space.
      This is both an advantage and a limitation. That is, from one hand a polynomial scaling allows 
      to reach large single-particle basis and medium mass nuclei, on the other hand the nuclear
      wave function has components in the full Hilbert space which grows exponentially in size with
      the size of the single-particle space. Presumably (or better hopefully) out of the full Hilbert space
      only a tiny fraction  gives the most important contributions to observables.
      The method we use, the Hybrid-Multi-determinant method (HMD) (ref.[5]), is rather different from the CC
       or IM-SRG, in the sense that no simple reference state is needed. We approximate the nuclear
      wave function as a linear combination of the most generic Slater determinants and 
      the coefficients of these Slater determinants, as well as the Slater determinants themselves, 
      are determined variationally using rank-3 
      gradient methods (ref.[6])(very similar to the well known BFGS method (ref.[7])). Also the HMD method uses a
      number of coefficients much smaller than the size of the Hilbert space. However
      analytically strongly founded extrapolation methods (refs. [8]-[10]) allow to estimate with some uncertainty
      the energy at zero energy variance (as it should  be for an eigenstate in the full Hilbert space).
      More precisely, suppose that we have an approximate eigenstate $|\psi>$ of the Hamiltonian,
      then the expectation value of the Hamiltonian is related to the energy variance obtained with
      this state by the relation  $<\HH>-E_{gs} = a < ( \HH - <\HH>)^2> $ , where $\HH$ is the many-body 
      Hamiltonian,  $a$ is a constant and $E_{gs}$ is the ground state energy in the full Hilbert space,
      provided the  state  $|\psi>$ is sufficiently close to the exact eigenstate. 
      A set of approximate wave functions would allow us to extract the ground state energy $E_{gs}$.
      This energy-variance-of-energy (EVE) method allows for a bridge  between a relatively
      small parametrization of the nuclear wave function and the full Hilbert space.
      We performed this extrapolation only
      for ${}^{24}Mg$ using $13$ major shells. This extrapolation is not necessary for the
      evaluation of the excitation energies, as described below. 
\par
 The HMD method is equally applicable to both closed shell and
 open shell nuclei. Although in this work we do not include a genuine
 NNN interaction, it is nonetheless interesting to see what predictions
 a reasonably soft NN interaction gives for excitation energies in the case
 of open shells nuclei, especially where collective behavior appears, without
 any renormalization.
\par
      As the NN interaction we consider the recently introduced chiral interaction by Entem, Machleidt and Nosyk
      (ref.[11]) without additional renormalization.
      The outline of this paper is as follows. In section 2 we briefly recap the HMD method.
      In section 3 we present the numerical results and in section 4 some conclusions and
      outlook.
\par
\section{ A brief recap of the the HMD method.}
\par
      The key idea of the HMD method is to expand the nuclear wave function as a linear
      combination of many generic Slater determinants (with exact or partial restoration
      of good quantum numbers using projectors) and to determine these Slater
      determinants using energy  minimization techniques. We use an harmonic oscillator basis.
      The wave-function of the nucleus is written as
$$
| \psi>= \sum_{S=1}^{N_D} g_S \HP |U_S>
\eqno(1)
$$
       where $\HP$ is a projector to good quantum numbers (e.g. good angular momentum and parity)
       $N_D$ is the number of Slater determinants $|U_S>$ expressed as
$$
|U_S> = \oc_1(S)\oc_2(S)... \oc_A(S) |0>.\;\; S=1,..,N_D
\eqno(2)
$$
       The generalized creation operators $\oc_{\alpha}(S)$ for $\al=1,2,..,A$ are a linear combination
       of the creation operators $\ad_i$ in the single-particle state labeled by $i$
$$
\oc_{\al}(S)=\sum_{i=1}^{N_s}U_{i,\al}(S)\ad_i  \;\;\;\;\; \al=1,...A
\eqno(3)
$$
       Here $N_s$ is the number of the single-particle states. These generalized creation operators
       depend on the Slater determinant $S$.
       The complex coefficients $U_{i,\al}(S)$ represent the single-particle wave-function of the
       particle $\al=1,2,..,A$. We do not impose any symmetry on the Slater determinants (axial or other)
       since the $U_{i,\al}(S)$  are variational parameters and good quantum numbers are restored using
       the projectors.
       These complex coefficients are obtained by minimizing the energy expectation values
$$
E[U]= { <\psi |\HH |\psi> \over <\psi |\psi>}
\eqno(4)
$$
       where $\HH$ is the total Hamiltonian, which also includes the usual center of mass
       Hamiltonian $\beta(\HH_{cm}-3/2 \hbar \om)$, in order to suppress excitations of the center of mass.
       The coefficients $g_S$ in eq. (1) are obtained by solving the generalized eigenvalue problem
$$
\sum_{S} <U_{S'} |\HP\HH | U_S> g_S = E \sum_{S} <U_{S'}|\HP| U_S> g_S
\eqno(5)
$$
       for the lowest eigenvalue $E$. 
      We have two versions of the method, which we call HMD-a and HMD-b.
      In the first version the two-body matrix elements of the Hamiltonian 
      $H_{1234}$ where 1,2,3,4 label the single-particle states with quantum numbers
      $1=(n_1,l_1,j_1,j_{z1},t_{z1})$, etc. ($n,l,j,j_z$ and $t_z$ denote the principal 
      quantum number, the orbital angular momentum, the angular momentum, its z-projection,
      and the isospin) all satisfy the relation $2 n+ l \leq e_{max}$. In the b-version
      the single-particle quantum numbers satisfy the relation 
      $2 n_1+l_1 +2 n_2+l_2 \leq N_{2max}$ (and similarly for the states 3 and 4).
      The b-version has been used by the author in the past only to test the variational programs
      (using renormalized interaction for the Deuteron binding energy an accuracy of one part
      in a million can easily be achieved). In this project we use bare interactions, that is
       no renormalization steps are performed. A renormalization of the two-body interaction 
      is necessary for strong interactions.
      The EMN interactions, especially at the $450 MeV$ cutoff seem to be soft
      enough so that we preferred to use bare interactions. This has the advantage that
      there are no induced many-body interactions, which are difficult to deal with.
      Presumably at large cutoff and medium mass nuclei a preliminary renormalization
      either with the Suzuki-Lee-Okamoto method (ref.[12] or the Similarity Renormalization Group seems
      advisable (ref.[13]). 
\par
In this work we use the HMD-a version for excitation energies. 
      The
      HMD-b version seems more convenient for binding energies since we can perform calculations with
      much larger single-particle states ($N_{2max}\simeq 13$). However, the HMD-b version seems to have a strong 
      dependence on the strength of the center of mass Hamiltonian $\beta$ and this feature
      has not been fully analyzed yet and it will not be discussed here. 
      Moreover for binding energies the final EVE  step 
      is necessary. This step is not necessary for excitation energies. The reason is the following.
      Consider for example the nucleus ${}^{24} Mg$ and the excitation energy of the first $2^+$
      state. We construct a sequence of approximate wave functions consisting of increasing numbers
      of Slater determinants $N_D$ and evaluate the the energy of the ground-state and of the first 
      $2^+_1$ state. The energies $ E_{gs}(N_D)$ and $ E_{2^+_1}(N_D)$ are not exact but
      they tend to the exact values as $N_D$ becomes larger and larger. That is,
      the exact energies would be $E_{gs}=E_{gs}(N_D)+ \delta_{gs}(N_D)$ and 
       $E_{2^+_1}=E_{2^+_1}(N_D)+ \delta_{2^+_1}(N_D)$. As $N_D $ goes to infinity the deltas
      tend to zero. The deltas are the errors in the two energies and have the same negative sign.
      When we take the difference in order to obtain the excitation energy these errors
      cancel out. Therefore for sufficiently large $N_D$ we should obtain excitation energies
      which have only a small dependence on $N_D$. Provided of course that we perform the variational
      calculations for both states exactly at the same level of approximation. Schematically these calculations
      start with $N_D=1$ (Hartree-Fock). We add a trial generic Salter determinant and minimize the
      energy expectation value  with respect to the last added Slater determinant.
      We call this  the "addition phase"). We then vary anew all
      Slater determinants  for  $D=1,2$ in sequence ("refinement phase") until the energy
      changes less than a termination value (typically $5\div 10 KeV$).
      We then keep adding new  Slater determinants. In the addition phase
      we vary only the one added last. After we reach a certain number of Slater determinants we
      repeat the refinement procedure to all Slater determinants until the termination criterion is met.
      The refinement phase is performed after we reach specified  numbers of Slater determinants
      typically after we reach $N_D=2,5,10,15,25,35,50,70,100,..$ (these numbers are simply a possible choice).
      Exactly the same procedure is implemented for the ground-state and for the excited states,
      since we want the approximate wave-functions for the ground-state and excited states to have the same
      degree of accuracy. Usually we use a partial $J_z^{\pi}$ projector to construct approximate wave functions.
      To restore the exact angular momentum quantum numbers we take the approximate wave functions with
      $N_D$ Slater determinants and reproject them to good $J^{\pi}$ in order to obtain better approximate
      excitation energies as a function of the number of Slater determinants $N_D$.

\bigskip
\section{ Numerical results.}
\bigskip
\par
       We focused mostly on four nuclei, $ {}^6Li, {}^{12}C$, ${}^{24}Mg$ and ${}^{48}Cr$.
      Experimental values for the excitation energies are from ref.[14]-[17] respectively
      (see also ref.[18]).
      Binding energies are from ref.[19]. In all calculations we considered single-particle states
      with $l <6$. All calculations use the N4LO-450 interaction.
      In all cases the harmonic oscillator frequency is selected around the minimum of
      the Hartree-Fock energy.
      In fig. 1 we show the dependence of the excitation energy of $3^+_1$ 
      state of ${}^6Li$
      as a function of the number of Slater determinants $N_D$. In this case the
      calculations have been performed at an harmonic oscillator frequency
      $\hbar\om=24MeV$. Note that the calculation does not include any coupling to the
      continuum. Experimentally the $3^+_1$ state is above in energy to the threshold
      of $\alpha+d$ break-up.
\begin{figure}
\centering
\includegraphics[width=10.0cm,height=10.0cm,angle=0]{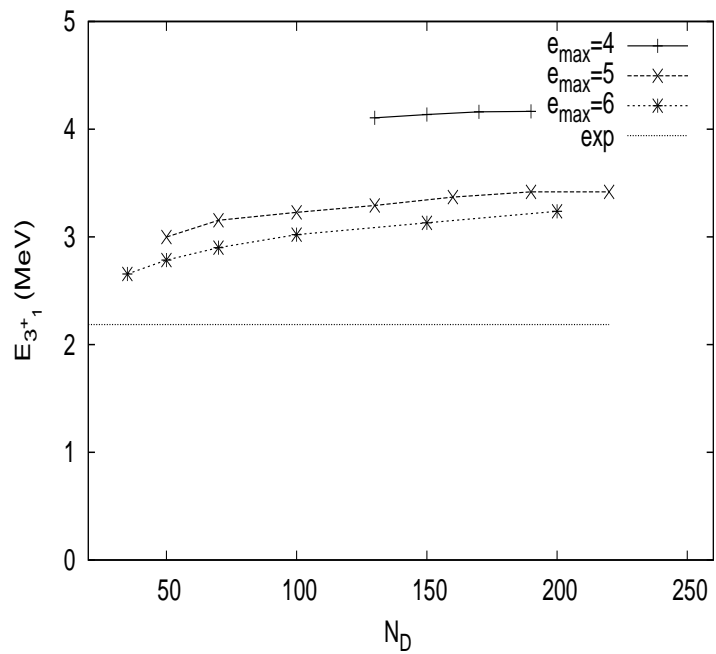}
\caption{ Excitation energy in MeV of the $3^+_1$ state of ${}^6Li$ as a function of
 the number of Slater determinants $N_D$  for several values of $e_{max}$
for the N4LO-450 interaction. The lines are only to guide the eye.}
\end{figure}
     In fig. 2 we show the behavior of the excitation energy of the $2^+_1$ of
     ${}^{12}C$ as a function of the number of Slater determinants. In this
     case we used an harmonic oscillator frequency of $\hbar\om=20MeV$.
\begin{figure}
\centering
\includegraphics[width=10.0cm,height=10.0cm,angle=0]{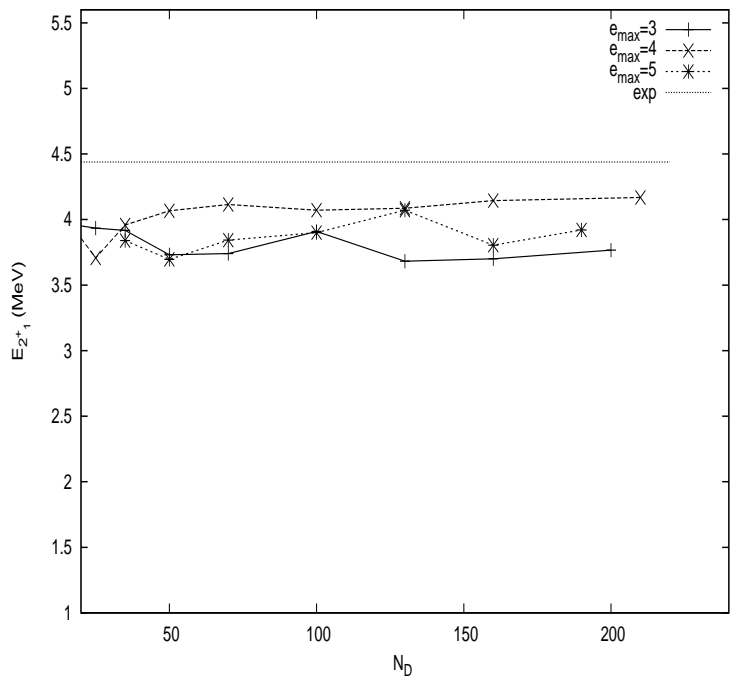}
\caption{ Excitation energy in MeV of the $2^+_1$ state of ${}^{12}C$ as a function of
 the number of Slater determinants $N_D$  for several values of $e_{max}$
for the N4LO-450 interaction.}
\end{figure}
    For these two cases the excitation energies are not too far off the experimental values.
    The nuclei ${}^{24}Mg$ and ${}^{48}Cr$ turned out to be the surprise. The excitation energy of the $2^+_1$
    state of ${}^{24}Mg$ is several times higher than the experimental one as shown in fig. 3.
    The experimental excitation energy of the $2^+_1$ state is $1.368 MeV$. In all evaluations of the 
    excitation energies, within a few hundred KeV's the convergence is reasonable, and it can be
    improved using more Slater determinants. The calculations have been performed at $\hbar\om=20 MeV$.
\begin{figure}
\centering
\includegraphics[width=10.0cm,height=10.0cm,angle=0]{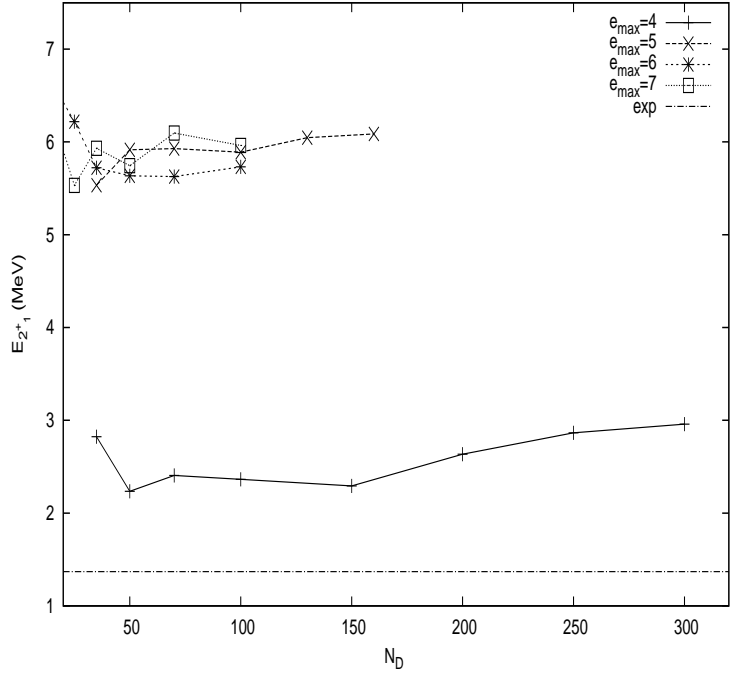}
\caption{ Excitation energy in MeV of the $2^+_1$ state of ${}^{24}Mg$ as a function of
 the number of Slater determinants $N_D$  for several values of $e_{max}$
for the N4LO-450 interaction.}
\end{figure}
    A similar result has been obtained for the doubly open shell nucleus ${}^{48}Cr$ as sown in fig. 4.
    The experimental excitation energy of the $2^+_1$ state is $0.752 MeV$
    In this case we used $\hbar\om=22 MeV$. Although we investigated very few cases it is striking 
    that for medium mass nuclei we obtain excitation energies too far off the experimental values.
\begin{figure}
\centering
\includegraphics[width=10.0cm,height=10.0cm,angle=0]{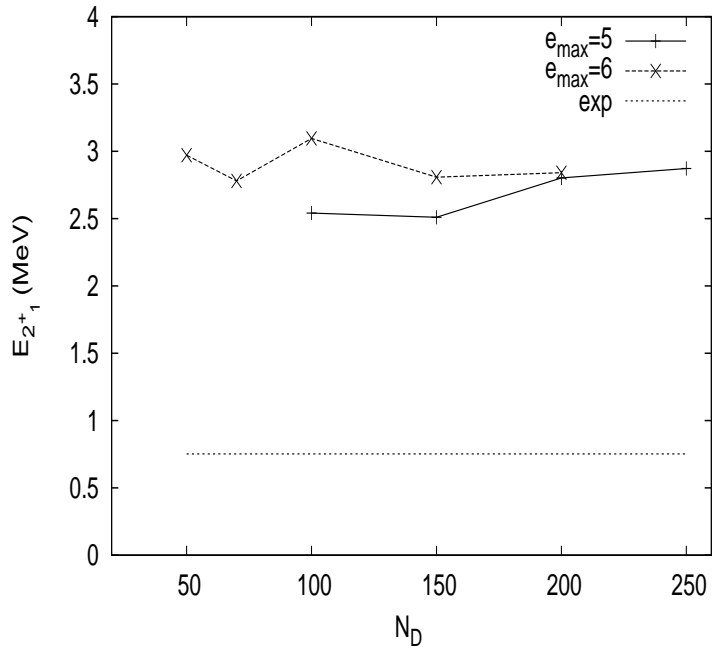}
\caption{ Excitation energy in MeV of the $2^+_1$ state of ${}^{48}Cr$ as a function of
 the number of Slater determinants $N_D$  for few values of $e_{max}$
for the N4LO-450 interaction.}
\end{figure}
     As mentioned in the introduction, we performed a EVE analysis only for ${}^{24}Mg$.
     For ${}^{48}Cr$ it was deemed unnecessary since even using only $25$ Slater determinants
     with $11$ major shells (both are small numbers) we reached the experimental binding energy.
     The interaction we used lacks the saturating effect of the NNN interaction and the NN interaction
     strongly overbinds. For ${}^{24}Mg$ we used $200$ optimized Slater determinants with $13$ major shells.
     The EVE analysis has been performed as follows. These $200$ Slater determinants $|U_S>$ with
     $S=1,..,200$,  were first determined
     without the use of angular momentum (partial or full) and parity projector. The minimization has been performed 
     as previously described.
Out of these $N_D=200$ Slater determinants we can form several
approximate nuclear wave functions. We could construct wave functions using
the first $1,2,..,n_S$ Slater determinants with $n_S=1,2,..$ up to $n_S=N_D$,  determine 
anew the coefficients of the linear combination using the Hill-Wheeler equations and determine
the variance of energy for these approximate nuclear wave functions. However only
for sufficiently large $n_S$ we have reasonably approximate wave functions. In practice
we evaluate the energy and the corresponding variance of energy for all $n_S=1,2,..,N_D$
and we keep only the points $(<\HH^2>-<\HH>^2,<\HH>)$ evaluated with reasonably accurate
wave functions (i.e. $n_S$ should be large enough) so that all points lie on a straight
line. Only then we fit the coefficients $a$ and $b$ in $E= a+b<(H-E)^2>$.
     The intercept $a$ is the estimate of the ground-state energy.
 The EVE plot is shown in fig.5.
     The final results for the coefficients $a$ and $b$ are $a= (-226.269  \pm 0.140)MeV$ and
     $b  = (0.01523 \pm 3.3\times 10^{-5})MeV^{-1}$.  The experimental binding energy is $198.256 MeV.$
\begin{figure}
\centering
\includegraphics[width=10.0cm,height=10.0cm,angle=0]{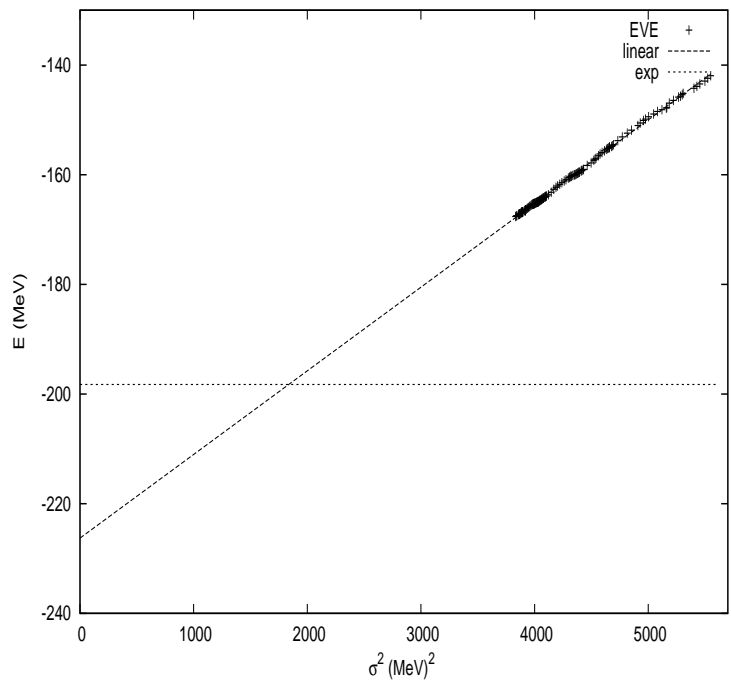}
\caption{ EVE plot for the ground-state of ${}^{24}Mg$ using $13$ major shells.
 The experimental value is shown as an horizontal line.}
\end{figure}
\bigskip
\section{ Conclusions and outlook.}
      In this work we considered the reasonably soft NN interaction N4LO-450 and performed some calculations
      about excitation energies away from major shell closure. In the cases of ${}^{24}Mg$ and ${}^{48}Cr$
      we did not obtain one of the typical features of collective behavior, i.e. low excitation energy.
      It could well be that the inclusion of the three-body interaction is necessary, a difficult task to implement.
      Another possible cause could be that our method of evaluating excitation energies must be pushed
       to a much larger number of Slater determinants. Or, a possible reason could be that the bare
      interaction couples too strongly low momentum and high momentum states. In other words, 
      a further renormalization must be used in order to obtain reasonable excitation energies.
      A renormalization procedure as done in SRG decouples low momentum from high momentum states. This
      can be tested with reasonable ease, and it will be the goal of future work.
\section{ Acknowledgments.}
      The author wishes to thank R.Machleidt for providing the EMN subroutines. Computational
      resources have been partially provided by a CINECA ISCRA-C project.
\vfill
\eject

\vfill
\eject

\bigskip
\begin{thebibliography}
\bigskip
\bibitem {1}
1. B. R. Barrett, P. Navrátil, and J. P. Vary, Prog. Part. Nucl. Phys.
69, 131 (2013).
\bibitem {2}
 G Hagen et al.  Rep. Prog. Phys. 77 096302(2014) 
\bibitem {3}
H. Hergert, S.K. Bogner, T.D. Morris, A. Schwenk, K. Tsukiyama.\\
Physics Reports 621 (2016) 165-222
\bibitem {4}
J. M. Yao, J. Engel, L. J. Wang,C. F. Jiao,\\
H. Hergert PhyS. Rev. C 98, 054311 (2018).
\bibitem {5}
 G.Puddu Eur. Phys. J. A 45, 233-238 (2010)\
 G. Puddu, J. Phys. G: Nucl. Part. Phys. 32, 321 (2006).\\
 G. Puddu, Eur. Phys. J. A 31, 163 (2007).\\
 G. Puddu, Eur. Phys. J. A 34, 413 (2007).
\bibitem {6}
 G.Puddu  Eur. Phys. J. A    42, 281(2009)
\bibitem {7}
  W. Lederman (Editor), Handbook of Applicable Mathematics, Vol. III,\\
 Numerical Methods (John Wiley and Sons, New York, 1981)
 Chapt. 11. and refs. in there
\bibitem {8}
Mizusaki T and Imada M  Phys. Rev. C 65 064319(2002)\\
Mizusaki T and Imada M  Phys. Rev. C 67 041301(2003)
\bibitem {9}
G.Puddu. J. Phys. G: Nucl. Part. Phys. 39 085108(2012).
\bibitem {10}
N. Shimizu et al.  Phys. Scr. 92 063001 (2017)
\bibitem {11}
D. R. Entem, R. Machleidt, and Y. Nosyk Phys. Rev. C 96, 024004(2017)
\bibitem {12}
K. Suzuki, S.Y. Lee, Prog. Theor. Phys. 64, 2091 (1980)\\
K. Suzuki, Prog. Theor. Phys. 68, 1627 (1982)\\
K. Suzuki, Prog. Theor. Phys. 68, 1999 (1982)\\
K. Suzuki, R. Okamoto, Prog. Theor. Phys. 92, 1045 (1994)
\bibitem {13}

S.K. Bogner a,b , R.J. Furnstahl c , A. Schwenk.\\
Progr. in Part. and Nucl. Phys. 65 (2010) 94.
\bibitem {14}
  Hu,Tilley,Kelley et al. Nucl. Physics A708, 3(2002).
\bibitem {15}
  R.B.Firestone. Nucl. Data Sheets 108, 2319(2007).
\bibitem {16}
  J.H.Kelley,J.E. Purcell and C.G.Sheu. Nucl. Physics A968, 71(2017).
\bibitem {17}
  T.W.Burrows. Nucl. Data Sheets 107, 1747(2006).
\bibitem {18}
https://www.nndc.bnl.gov/nudat2/
\bibitem {19}
G. Audi and A.H. Wapstra.  Nucl. Phys. A565, 1 (1993).
\end{thebibliography}
\end{document}